%Paper: hep-th/9505176
%From: HSLA@mitlns.mit.edu
%Date: Mon, 29 May 1995 20:32:41 -0400 (EDT)
%Date (revised): Mon, 29 May 1995 20:53:28 -0400 (EDT)

\input harvmac

%\font\caps=cmcsc10

\def\hat{\widehat}
\def\tilde{\widetilde}

\def\det{{\rm det}}

 %\lfr12: small \hf  or use {\textstyle     }
\def\half{{\textstyle{1\over 2}}}
\def\lfr#1#2{{\textstyle{#1\over#2}}} %       little fraction
\def\d{{\rm d}}
\def\e{{\rm e}}
\def\pa{\partial}
\def\mbox#1#2{\vcenter{\hrule \hbox{\vrule height#2in
		\kern#1in \vrule} \hrule}}  %e.g. \mbox{.1}{.1}

\font\cmss=cmss10 \font\cmsss=cmss10 scaled 833%at 7pt
\def\IZ{\relax\ifmmode\mathchoice
{\hbox{\cmss Z\kern-.4em Z}}{\hbox{\cmss Z\kern-.4em Z}}
{\lower.9pt\hbox{\cmsss Z\kern-.4em Z}}
{\lower1.2pt\hbox{\cmsss Z\kern-.4em Z}}\else{\cmss Z\kern-.4em Z}\fi}

\def\tr{{\rm tr}}

\def\CS{{\cal S}}
\def\CO{{\cal O}}

\def\mn{\mu\nu}

\noblackbox

\Title{\vbox{\baselineskip12pt{\hbox{MIT-CTP-2435}}%
{\hbox{hep-th/9505176}}}}
{\vbox{\centerline{Massive Dilaton and Topological Gravity }}}

\centerline{HoSeong ~La\footnote{$^*$}{%and Someone else{%
e-mail address: hsla@mitlns.mit.edu}   }

\bigskip
\centerline{Center for Theoretical Physics}
\centerline{Laboratory for Nuclear Science}
\centerline{Massachusetts Institute of Technology}
\centerline{Cambridge, MA 02139-4307, USA}
\vskip 0.7in

A model in which the massive dilaton is introduced by minimally extending
the two dimensional topological gravity is studied semi-classically.
The theory is no longer topological because of the explicit Weyl scale
symmetry breaking.
Due to the dilaton the semiclassical stress-energy tensor gets renormalized
and it is shown how the gravitational background coupled to the
the dilaton depends on the dilaton mass as well as the renormalization
mass scale, but not on the Newton's constant.

\bigbreak\bigskip\bigskip
\centerline{Submitted to {\it Physics Letters B}}
\Date{05/95} %replace this line by \draft  for preliminary versions
%\draft%\nolabels                           %%% later delete this
\noblackbox

\vfill\eject

\newsec{Introduction}

The two-dimensional gravity defined by the Einstein-Hilbert action
is trivial at the classical level in the sense that any two-dimensional
metric satisfies the classical equation of motion. No cosmological constant is
allowed at the classical level unless matter is present.
Quantization, nevertheless, reveals fairly
nontrivial structures of the theory, although it is not yet dynamical%
\ref\wittg{E. Witten, ``Topological Gravity,'' Phys. Lett. {\bf 206B} (1988)
601.}%
\ref\lpwtg{J. Labastida, M. Pernici and E. Witten, ``Topological Gravity in
Two Dimensions,'' Nucl. Phys. {\bf B310} (1988) 611\semi
D. Montano and J. Sonnenschein, ``Topological Strings,'' Nucl.
Phys. {\bf B313} (1989) 258.}.
There are nontrivial classical observables as topological invariants.
Being topological, such a theory in its essence is far from the real nature,
even in four dimensions.
One simple reason we can think about is that we live in the world
in which the scale symmetry is broken, but topological gravity respects
the scale symmetry.

In this letter we intend to investigate what might happen to the
two-dimensional topological gravity if we begin to break the topological
structure by hand, keeping in mind that we can take an analogous approach to
the four-dimensional topological gravity in the future. It would be much better
if we could find a mechanism to break the topological structure in itself,
nevertheless we expect the system should still evolve along a similar path to
reach to the broken phase.

The minimal thing we shall do is to break the scale symmetry,
but leave the theory still reparametrization invariant. One way to achieve this
is to introduce a real massive scalar field coupled to the gravity without
self-interactions. The case with self-interactions is in progress and
will be reported elsewhere\ref\intdil{H.S. La, in preparation.}.
This real scalar field can be identified as the
dilaton because its presence breaks the scale symmetry explicitly through the
mass term.
Note that in two dimensions there is no spontaneous symmetry breaking%
\ref\cole{S. Coleman, Comm. Math. Phys. {\bf 31}(1973) 259.}
to identify the dilaton as Goldstone boson so that this naming
is nothing illegal.
On the contrary to the massless case, there is no classical level conformal
symmetry to begin with and
the stress-energy tensor is not traceless due to the mass of
the dilaton unless the equation of motion is imposed.

As a result, although the graviton is not dynamical yet (i.e. not
propagating), the theory has nontrivial dynamical degrees of freedom due to the
dilaton. Full quantization will involve a Liouville-like mode (which is not
exactly Liouville because of different potential form, but is induced from
the conformal factor of the metric), however it
is no longer regarded as an anomaly. It is just a quantum correction to the
stress-energy tensor. Since the quantization of such a mode is still
elusive, under this circumstance we are tempted to investigate this theory
using semi-classical method as a first step to see what happens.
(Later in the discussion, the rationale behind such an approach will be
 addressed more.)
In other words, a quantum field theory of the dilaton in a classical
gravitational background will be studied.
In principle, we can work in any gravitational
background, but we shall focus on the de Sitter background for an example.
Other cases will be reported in \intdil.

\newsec{Lagrangian}

The model we are interested in is described by the following action:
\eqn\elag{\CS=\CS_{{\rm top}} + \CS_{{\rm d}}
	=\int_\Sigma\! d^2x \left(\CL_{{\rm top}} +\CL_{{\rm d}}\right),}
where
\eqna\etop
$$\eqalignno{
\CL_{{\rm top}} &={\sqrt{-g}\over 16\pi G_0} R, 	&\etop a\cr
\noalign{\smallskip\smallskip}
\CL_{{\rm d}} &=-\sqrt{-g}\left(\half g^{\mn}\pa_\mu\phi \pa_\nu\phi
 	+\half m^2\phi^2\right). &\etop b\cr
}$$
We have chosen the metric signature $(-+)$, but the Euclidean case
can be easily obtained by simply replacing all the minus signs in the
above with the plus sign.
The first term in eq.\elag\ is a constant (Euler number) to define the
classical
topological gravity and the second term introduces dynamics to the system.
Without the mass term in eq.\etop{b}, the action is invariant under Weyl
rescaling $g_{\mn}\to \e^{2\rho}g_{\mn}$ which is the two dimensional
analogue of the scale symmetry.

The stress-energy tensor is given by
\eqn\eten{T_{\mu\nu}=-{2\over\sqrt{-g}}{\delta\CS_{{\rm d}}
\over\delta g^{\mu\nu}}=\pa_\mu\phi\pa_\nu\phi
- \half g_{\mn}g^{\alpha\beta}\pa_\alpha\phi\pa_\beta\phi
- \half g_{\mn}m^2\phi^2}
such that the trace is
\eqn\ei{T_{\ \mu}^{\mu}= - m^2\phi^2.}
Thus the dilaton mass breaks the Weyl symmetry.
Note that in two dimensions the existence of
a massless real scalar field does not break the scale symmetry,
but the mass term does.
This, however, does not necessarily imply that the classical vacuum of
eq.\elag\ does not respect the conformal symmetry.
The classical equation of motion of eq.\elag\ under the variation of
the metric is nothing but $T_{\mn}=0$
using the property of two dimensional geometry
$R_{\mn}=\half g_{\mn} R$. Together with eq.\ei, $T_{\ \mu}^{\mu}=0$ and then
$\phi=0$ is the only classically allowed value so that the
dilaton does not appear explicitly. In other words, classically
$\CS_{{\rm d}}=0$ and $\CS$ is still a topologically invariant quantity
to be the Lagrangian of topological gravity.

The classical equation of motion of the dilaton is
\eqn\eomdil{(\Delta+m^2)\,\phi=0,}
where $\Delta\equiv -{1\over\sqrt{-g}}\pa_\mu\sqrt{-g} g^{\mn}\pa^\nu$ is
the Laplacian. Note that $\phi=0$ is a trivial solution and other solutions
fail to satisfy $T_{\mn}=0$.

One can add a cosmological constant
$\CL_{\rm c}=-\sqrt{-g}\Lambda_0/8\pi G_0$
that also breaks the Weyl symmetry.
In this case the classical equation of motion modifies to read
$8\pi G_0 T_{\mn} =g_{\mn}\Lambda_0$, whose solution
$4\pi G_0 m^2\phi^2 = -\Lambda_0 $ still leads to $\CL_{\rm c} +\CL_\d  =0$.
Thus there is no consequence of the classical
dilaton. However the situation becomes much different quantum mechanically.

\newsec{Quantum Physics of Dilaton}

As alluded in the introduction,
 we shall incorporate the gravity classically and
investigate the implications of quantum physics of the dilaton. Although
the validity of such a semi-classical approach may be always arguable,
for our purpose it should be still good enough to learn the quantum
effect of the dilaton near the classical topological vacuum.
We use the functional integral method to compute the quantum effect of the
dilaton.
Thus integrating out the dilaton, we obtain
\eqn\eii{W_\d \left(\equiv\int\! d^2x\CL_{{\rm d,eff}}\right)
={\textstyle {i\over 2}} \tr\ln(\Delta +m^2).
}
If the exact result of the right-hand side were known,
the semi-classical equation could be derived by simply taking the
variation with respect to the metric.
Unfortunately, the exact result is unknown so that we have to take a detour.
First, we shall compute the divergent contributions to renormalize
$\CS_{{\rm top}}$, then the rest will be computed to derive the
semi-classical equation.

For a real scalar field this has been computed using the WKB approximation
method in
\ref\dewitt{B.S. DeWitt, ``Quantum Field Theory in Curved Spacetime,''
Phys. Rep. {\bf 19} (1975) 295.}%
\ref\birdav{ N.D. Birrell and P.C.W. Davies,
{\it Quantum Fields in Curved Space}, (Cambridge, 1982).}
so that we can just recapture the result in two dimensions.
We expect that there would be ultraviolet divergences and it is important to
isolate them to renormalize. We follow DeWitt who used Schwinger's method
to handle divergent terms%
\ref\schwin{J.S. Schwinger, Phys. Rev. {\bf 82} (1951) 664.}%
\dewitt.
The essence of this method involves representing the propagator as
\eqn\eiii{{1\over \Delta + m^2}=i\int_0^\infty\!
d\tau\,\e^{-i\tau(\Delta + m^2)}.}
Then we compute $W_\d$ for $d=2+\epsilon$
and use the dimensional regularization to take $\epsilon\to 0$ limit later:
\eqna\eiv
$$\eqalignno{W_\d\! &=\!{\textstyle {i\over 2}}\int\! d^dx\sqrt{-g}\,
\langle x|\ln(\Delta+m^2)|x\rangle &\eiv a\cr
&=\!\half\int\! d^dx\sqrt{-g}
\lim_{x'\to x}\int_0^\infty \!\!\! d\tau\! \int_{m^2}^\infty\! dm^2
\langle x|\e^{-i\tau(\Delta + m^2)}|x'\rangle  	&\eiv b\cr
&=\!{\textstyle {i\over 2}}\int\! d^dx\sqrt{-g}
\lim_{x'\to x}\!\int_0^\infty\!\!\! d\tau \!\int_{m^2}^\infty\! dm^2
{1\over (4\pi i\tau)^{d/2}}\left[V(x,x')\right]^{1/2}
\e^{-i\left(\tau m^2-{\sigma\over 2\tau}\right)}F(x,x';i\tau).
 	&\cr
& &\eiv c\cr
}
$$
In the above equation, we use $\sigma$ which is one-half of the square of the
proper distance between $x$ and $x'$
\eqn\ev{\sigma(x,x')=\half y_\alpha y^\alpha}
in terms of the Riemann normal coordinates $y^\alpha$ of $x$ with origin at
$x'$, and the Van Vleck determinant is defined by
\eqn\evanvl{V(x,x')\equiv -{1\over \sqrt{g(x) g(x')}}\,
\det\left(\pa_\mu\pa_\nu \sigma(x,x')\right).}

Since we are interested in the short distance behavior of $F$,
assuming that the high frequency behavior of the massive scalar field is
relatively insensitive to the long term time-dependence of the metric we
study, we introduce the following asymptotic adiabatic expansion
\eqn\evi{F(x,x';i\tau)=a_0(x,x') + a_1(x,x')i\tau + a_2(x,x')(i\tau)^2 +\cdots}
In this approximation eq.\eiv{c}\ is easily computed to yield
\eqn\evii{\CL_{{\rm d,eff}}={\sqrt{-g}\over 2(4\pi)^{d/2}}
\left({m\over\mu}\right)^{d-2}\sum_{j=0}^\infty
a_j(x)(m^2)^{1-j}\Gamma(j-{\textstyle {d\over 2}}),}
where $a_0=1$, $a_1=R/6$
and the regularization mass scale $\mu$ is introduced explicitly.
Using the infinitesimal property of the Gamma function
\eqn\eviii{\Gamma(-\lfr{\epsilon}{2})=-{2\over \epsilon} - \gamma
+\CO(\epsilon),\quad \gamma={\rm Euler \ constant},}
the ultraviolet divergences of the effective Lagrangian can be identified as
\eqn\eix{\CL_{{\rm d,eff}}= {\sqrt{-g}\over 8\pi}
\left[m^2\left(N_\epsilon + \ln{m^2\over \mu^2} -1 \right)
-{R\over 6}\left(N_\epsilon + \ln{m^2\over \mu^2} \right)
\right]+\CO(\epsilon)+\cdots}
where $N_\epsilon\equiv {2/ \epsilon} +\gamma -\ln 4\pi$.
The ellipsis includes higher order finite corrections to the effective
Lagrangian, which does not necessarily vanish in $\epsilon\to 0$ limit.

Renormalizing these divergences corresponds to dilatonic corrections to the
cosmological constant and the Newton's constant
(the coefficient of the order $R$ term) in the effective action.
Thus in this case we obtain renormalized semi-classical equation
$${R_{\mn}-\half g_{\mn}R +  g_{\mn} \Lambda_{{\rm ren}}
= 8\pi G_{{\rm ren}}\langle T_{\mn}\rangle_{{\rm ren}}}$$
where
\eqna\ecosc
$$\eqalignno{{1\over 16 \pi G_{{\rm ren}}}&={1\over 16\pi G_0}
-{1\over 48\pi}\left(\ln{m^2\over \mu^2}+N_\epsilon \right)\ \ ,
						&\ecosc a\cr
\Lambda_{{\rm ren}}&
=G_{{\rm ren}}\left({\Lambda_0\over G_0}
- m^2\left(\ln{m^2\over\mu^2}+N_\epsilon -1\right)\right).  &\ecosc b\cr
}$$
Note that we have introduced the bare cosmological constant $\Lambda_0$
to renormalize the functional integral.
$\Lambda_{{\rm ren}}$ cannot be made finite without $\Lambda_0$.

$\langle T_{\mn}\rangle_{{\rm ren}}$ on the right-hand side collects the
rest of finite quantum contributions of the massive dilaton to
the effective action, which should be computed independently. The adiabatic
expansion eq.\evi\ is only useful for computing divergences so that we still
need to take care of other quantum corrections.
The above are renormalization scheme independent because the
ambiguity of the subtraction scheme in introducing counter terms
are absorbed into the mass scale $\mu$ by proper redefinition. Such
$G_{{\rm ren}},\ \Lambda_{{\rm ren}}$ and $\mu$ are
usually determined by measurement.

We can immediately observe the follows: a heavy dilaton $(m^2>\mu^2)$
increases the Newton's constant, whilst a light dilaton $(m^2<\mu^2)$
decreases. In four dimensions the Newton's constant controls the strength
of the gravity, but here, as we shall find out soon,
the effective gravitational equation does not depend on the
renormalized Newton's constant explicitly to confirm that the gravity is
not dynamical.
The curvature generated will
depend solely on the dilaton mass and the regularization mass scale.
For some $m^2$ the dilaton-corrected cosmological constant
$\Lambda_{{\rm ren}}$ could vanish. Without loss of generality we assume
$\Lambda_{{\rm ren}}=0$ at $\ln(m^2/\mu^2) = 0.$ For this we have
 $\Lambda_0/ G_0=m^2 (N_\epsilon -1)$ to obtain
\eqn\erfx{\Lambda_{{\rm ren}} = -G_{{\rm ren}}m^2\ln{m^2\over \mu^2}.}
Then  $\Lambda_{{\rm ren}}$ is negative for large $m^2$,
whilst for small $m^2$, $\Lambda_{{\rm ren}}$ is positive.
It may look like that there is no cosmological constant
generation by a massless dilaton, but
this is not quite true. Eq.\erfx\ only tells us that there is no ultraviolet
contribution to the cosmological constant from a massless dilaton. In the
massless dilaton case, eq.\eii\ has infrared divergences, which can be a source
to generation of a cosmological constant%
\ref\polystr{A.M. Polyakov, Phys. Lett. {\bf 103B} (1981) 207\semi
A.M. Polyakov, ``{\it Gauge Fields and Strings},'' (Harwood Academic, 1987).}.
A massive dilaton automatically removes such an infrared divergence.

In principle we can add higher order finite correction terms to the
semi-classical equation, e.g. a curvature square term,
we shall however focus only on the leading correction in this letter.
In two dimensions every metric satisfies $R_{\mn}=\half g_{\mn}R$ the
semi-classical gravitational equation now becomes nothing but
\eqn\esemi{ g_{\mn} \Lambda_{{\rm ren}} = 8\pi G_{{\rm ren}}
\langle T_{\mn}\rangle_{{\rm ren}}.}
Using eq.\erfx, we find that this effective equation indeed
does not depend on the renormalized Newton's constant $G_{{\rm ren}}$.

We now need to compute $\langle T_{\mn}\rangle_{{\rm ren}}$ which is the
renormalized piece of $\langle T_{\mn}\rangle$. To be consistent we shall use
dimensional regularization, but
in general it is not easy to compute this expectation value
$\langle T_{\mn}\rangle$ in dimensional regularization.
However in the de Sitter space the situation is better because
\eqn\exi{\langle T_{\mn}\rangle = \half g_{\mn}\langle
		T_{\alpha}^{\ \alpha}\rangle
	=-\half g_{\mn} m^2 \langle\phi^2\rangle.}
$\langle\phi^2\rangle$ can be computed using the result in
\ref\canrai{P. Candelas and D.J. Raine, Phys. Rev. D {\bf 12} (1975) 965.}
or \birdav\ such that
\eqn\exii{
\eqalign{\langle\phi^2\rangle &= {2\over R}\left({R\over 8\pi}\right)^{d/2}
{\Gamma\left(\nu(d)-\half +{d\over 2}\right)
\Gamma\left(-\nu(d)-\half +{d\over 2}\right)
\over \Gamma\left(\half +\nu(d)\right)\Gamma\left(\half -\nu(d)\right)}
\Gamma\left(1-{\textstyle {d\over 2}}\right)\cr
&={1\over 4\pi}\left( -{2\over\epsilon}-\gamma +\ln 4\pi\right)
+ {1\over 4\pi}\left(\psi(\nu_+) + \psi(\nu_-) +\ln{2\mu^2\over R}\right)
+\CO(\epsilon) \cr}}
where $\nu_\pm \equiv\half\pm\nu,\ \nu^2\equiv{1\over 4}-{2m^2\over R}$,
$(\nu(d))^2= {1\over 4}(d-1)^2 -{d(d-1)m^2\over R}$
and $\psi$ is the Digamma function. Note that the same mass scale $\mu$ is
used for $R$. Using the
same renormalization scheme we used for $\CL_{{\rm d, eff}}$, we can obtain
\eqn\exiii{\langle\phi^2\rangle_{{\rm ren}}
=\langle\phi^2\rangle - \langle\phi^2\rangle_{{\rm adiabatic}}
= {1\over 4\pi}\left(
\psi(\nu_+) +\psi(\nu_-) +\ln {2\mu^2\over R}-{R\over 6m^2}+\ln{m^2\over\mu^2}
\right)
}
where $\langle\phi^2\rangle_{{\rm adiabatic}}$ that takes care of the
counter term contributions can be easily computed from
eq.\eiv{c} using the definition
$\CL_{{\rm d, eff}} = {i\over 2}\int_{m^2}^\infty
dm^2\langle\phi^2\rangle_{{\rm adiabatic}}$.
Then we obtain
\eqn\exiv{\langle T_{\mn}\rangle_{{\rm ren}} = -{1\over 8\pi}
g_{\mn}\left[m^2\left(\psi(\nu_+) + \psi(\nu_-) +\ln 2\right)
+m^2\ln{m^2\over\mu^2} -m^2\ln{R\over\mu^2} - {1\over 6}R\right].}
Note that as $m^2\to 0$, $\langle T_{\mn}\rangle_{{\rm ren}}\to {1\over 48\pi}
g_{\mn}R$ to correctly recover the conformal anomaly in the massless case.

Eq.\esemi\ now becomes
\eqn\eceq{0 =\ln{R\over\tilde{\mu}^2} -\psi(\nu_+) - \psi(\nu_-)
+{1\over 6m^2} R}
where $\ln\tilde{\mu}^2 \equiv \ln2\mu^2$.
This is the key equation of this letter.
Solutions to this equation can be obtained as the intersection
points  of the following two curves:
\eqna\exv
$$\eqalignno{f_1 (\hat{R}) &=\psi(\nu_+) +\psi(\nu_-) -{1\over 6} \hat{R}
&\exv a\cr
f_2 (\hat{R}; \hat\mu^2) &=\ln {\hat{R}\over\hat{\mu}^2} &\exv b\cr
}$$
where $\hat{R}\equiv R/m^2$ and $\hat\mu\equiv\tilde\mu/m$.
Since $f_1$ is a decreasing function from $\lim_{R\to 0}f_1\to 0$ to
$\lim_{R\to\infty} f_1\to -\infty$, there is always an intersection with $f_2$.
Note that the intersection points depend only on $m^2$ and $\mu^2$.

Let us call the curvature at the intersection point
$R_d(m^2, \tilde\mu^2)$, whose exact analytic form is not really important
at this moment. $\hat R_d$ increases as $\hat\mu^2$ increases and
$f_1<0$ implies that $\hat R_\d<\hat\mu^2$. The above relations work
regardlessly whether $\nu$ is real or imaginary because
${\rm Im} \left(\psi(\nu_+) +\psi(\nu_-)\right)$ always vanishes.
It nevertheless constrains $\hat R_\d \geq (<)8$
if $\ln\hat\mu^2\geq(<)\  4/3 +2\gamma +7\ln 2$
respectively.
In principle, $\tilde\mu^2$ is independent from $m^2$ so that $\hat R$
can take any value according to the value of $\hat\mu$.
However, $\tilde\mu^2$ should be of order $m^2$ because
$m^2$ is the only scale of our model to start with and there is no
reason to have a completely different mass scale to generate any
hierarchical mystery.
Thus $\hat\mu \sim \CO(1)$ for reasonable solutions. Thus $R\sim\CO(m^2)$
and in particular $R$ vanishes for the massless dilaton in this limit
(this can be easily shown from eq.\ecosc{b}\ and eq.\esemi.).

\newsec{Conclusion}

We have seen so far that the massive dilaton explicitly breaks
the Weyl symmetry so that the curvature of the gravitational background
coupled to the dilaton is no longer arbitrary but constrained by
the mass of the dilaton at least in the de Sitter space.
In particular we have seen that the curvature does not depend on the Newton's
constant.
As a result, the two-dimensional gravity is not trivial any more.

It is certainly arguable if eq.\eceq\ is a consistent equation to address the
current issue.\foot{Some issues on the validity of the semi-classical
approximation in quantum gravity is discussed in
\ref\samir{G.Lifsdhytz, S. Mathur and M. Ortiz, ``A Note on the Semi-Classical
Approximation in Quantum Gravity,'' MIT-CTP-2384 (gr-qc/9412040).}.
}
One should first worry about that quantization of dilaton
without quantizing gravity is indeed a reasonable thing to do.
The dilaton is usually
regarded as part of the gravity rather than a matter field, but in two
dimensions we could treat it as a matter field. This is possible because
$\CL_{{\rm top}}$ in eq.\etop{a}\ is invariant under the Weyl rescaling so that
we cannot generate the dilaton field using the conformal property of the
curvature as in the four dimensions.
Due to the absence of Goldstone boson we could simply identify a real massive
scalar field that breaks the scale symmetry as the dilaton.
We avoided the quantization of the gravity simply because we do not know
how to quantize the Liouville-like mode. Even though we did that,
but, then, it would be difficult to separate and investigate
the quantum effect of the dilaton to the system.
Thus the semi-classical approach we took in this letter certainly serves our
purpose.
After knowing the fact that the dilaton
constrains the scale of the curvature through its mass, we can fully
quantize to investigate what kind of dynamical two-dimensional gravity
can be obtained as the broken phase of the two-dimensional topological gravity.

We certainly do not know how dilaton becomes massive, needless to say, neither
do we know how two-dimensional topological gravity generates the dilaton.
Presumably it would be better off to address these questions in a theory
where an extra symmetry related to the dilaton is present,
e.g. $N=2$ supergravity that becomes topological by twisting%
\ref\ntwotop{E. Witten, ``On the Structure of the Topological Phase of Two
Dimensional Gravity,'' Nucl. Phys. {\bf B340} (1990) 281.}.
Anyhow, once a massive dilaton is obtained, its quantum effect
should not be too much different from the result given here.

It would be interesting to see a similar effect in the four-dimensional
topological gravity\wittg\ by properly introducing a massive dilaton.
In such a way, perhaps we could unlock the mystery of the relationship
between the broken phase and the unbroken phase of topological gravity.

%\bigskip\bigskip\centerline{{\bf Acknowledgements}}
%\medskip
%The author thanks

%\appendix{A}{}
%{\leftline{\bf Note Added:}}

%
\listrefs
\vfill\eject
%\leftline{\bf Figure caption:}          \bigskip
%Fig. 1: Contributions needed to make the 3-loop diagram well-defined.
\bye